# Exploring the Research Landscape of Pakistan: A Data-driven Analysis of Scopus Indexed Scientific Literature

Muhammad Bilal Ibrahim [a], Saeed-Ul Hassan [a]

[a] Information Technology University, 346-B, Ferozepur Road, Lahore, Pakistan

## Abstract

Scientific contribution and research performance of a university, research group, or institute needs to be evaluated all the more with the increasing volume and fast-developing disciplines of research. The need of the time is to develop tools for strategic planning and management that will help research bodies to rank and benchmark themselves against international standards. This will enable them to invest appropriately in research areas of promising strength and gain maximally from them, thus fulfilling the ultimate purpose of positive impact of research on society. Our tool is capable of rating and benchmarking universities as well as research institutes in not only the major disciplines and sub-disciplines, but at the finest level of niche areas of science and technology too with the help of its innovative bibliometric indicators based on publications and citation analysis. The tool accepts inputs like discipline/subject area, university, and country and time window while using data retrieved from bibliography database, Scopus, to benchmark and rate the research body under consideration. We have evaluated that there are many niche subject areas in which small or medium size universities are performing good in comparison to the large universities. Most of these subject areas are of more significance in the present day and the future. Government and funds allocating bodies should take this factor in account that investing the right money at right place will give far better results than we they are having right now.

## Introduction

With the rapid increase in population, citizens and governments are working together for a rise in the living standards of people. A steep incline in the rate of education in developing countries has enhanced awareness about research among laymen. The knowledge gained through research and development is transferred via teaching, learning, and training to the coming generations, hence making ground for creating innovative options of research [1]. Education at universities is the next level of learning process, which functions as a critical component of human development all over the world [2]. With the diversity of students, recruitment of high-quality staff, as well as a limitation of resources in universities and research institutions in developing countries, it becomes crucial for them to maintain quality standards in research. Good quality research is capable of enhancing individual intellectual as well as creative capacity in nations;

this, in turn, enables nations to adopt and assess available research paths followed by pursuing and implementing those most suitable to their own resources and manpower. The need of the time is to develop tools for strategic planning and management that will help research bodies to rank and benchmark themselves against international standards. This will enable them to invest appropriately in research areas of promising strength and gain maximally from them, thus fulfilling the ultimate purpose of positive impact of research on society.

Research activity in an institute or university is reflected by the production of scientific literature. "*Publish or perish*" is a fact of research and academia, which warns researchers of the expected downfall of the value of their research if they do not present it to relevant professionals and researchers [3]. This means that a group of researchers may conduct research significant in size and content, but may fail to prove themselves worthwhile if they do not present it to the scientific community through available modes of publishing like high-quality peer-reviewed academic journals and conference papers. The literature published by a research body can be used to estimate the volume and quality of research by quantitative analysis of number of publications produced over time and their citation by other workers in the field, respectively [4]. The former is an indication of any group's scientific activities while the citation of these publications is an indication of scientific influence and impact [5]. With time, a research body like a university may increase its enrollment or a country may set up new universities and hence release more published literature i.e., "expand out" [6]. If the research work/education is of high quality, it will result in "expanding up" of its research as well. It is stated that for quality higher education, both "expanding up" and "expanding out" are necessary to cater the fast growing number of students.

Scientific contribution and research performance of a university, research group, or institute needs to be evaluated all the more with the increasing volume and fast-developing disciplines of research. According to Humphrey and Lukka, "Even in today's measurement obsessed world, there are still many things that count but that do not get measured  - and many things that are measured but not recognized or treated properly" [7]. Until all research performance and activities are calculated for quality and quantity, a useful balance between expanding up and expanding out cannot be attained. Until relatively recently, the major mode through which research was evaluated was peer-review or peer-judgement, in which expert personnel reviewed and gave their opinion about the worth of research under consideration [8]. Bibliometrics, on the other hand, is the statistical or quantitative analysis of all published products of research as well as their citation counts with the purpose of evaluating the health and progress of educational or research systems [9, 10]. In contrast to peer-review, it uses weighted quantitative indicators to measure the quality of research with homogeneity and reliability. Many technologies have been developed for rating and benchmarking of journals by publishing corporations in order to aid research agencies, university leaders, and academicians to organize, measure, and record research in all disciplines [11]. The challenge is to apply bibliometric indicators, such as number of research publications produced and citations of publications, to benchmark and rate research performance of research institutes and universities in developing countries to enable a comparison between publication-producing entities in a range of disciplines and sub-disciplines of science and technology.

With a mode of measurement of research, research bodies can be ranked or rated by ordering them as per their performance and/or through other indicators like input of resources, amount of research output, and impact of that research [1]. Universities are judged across all the core missions that are laid out for them, including teaching, research, knowledge transfer, and international outlook [12]. This process is called rating whereby performance is measured against predetermined indicators and is also meant to assess the value or worth of research activities taking place at a university or organization in terms of their social and economic impact [13]. This rating forms the basis of a systematic process of benchmarking in which research bodies are compared with respect to their performance against the ones with best ranks. Benchmarking is a comparative process in which any working body like organizations evaluate various aspects of their functions in comparison to the best practice within their own sector [13]. This is a prerequisite for research organizations and universities to develop plans in order to bring them at par with the best practitioners in their field.

Unfortunately, quality is not a priority in higher education or research scenario in developing countries like Pakistan, where due to limited facilities and lack of quality assurance mechanisms, the quality of research and education is deteriorating rapidly [19]. Here, there is a lack of an internationally competitive tool to measure quality of research so as to implement suitable quality assurance mechanisms. In 2003, Higher Education Commission, Pakistan, developed a Quality Assurance Committee meant to design a framework for accreditation and ranking of universities followed by development of a Quality Assurance Agency in 2005, both of which worked by arrangement of resources and seminars/workshops to enhance capacity building [20]; but an accurate tool to measure the improvement required and in the disciplines, sub-disciplines, or niche areas of concern was entirely absent. There are deficiencies in skills and research outputs of Pakistan that is why only a few of our universities rank in the top 500 of the world's universities [21]. There is a requirement of a tool, which can help leaders assess the quality of research performed in Pakistan and other developing countries at levels as fine as the niche areas, which will help them locate their strengths and allocate investments in the areas of development so as to enhance the quality of research.

Computational tools are capable of enabling research bodies to assess their research performance and know their areas of expertise in comparison to other institutes and universities. However, the current tools are designed to cater only a limited number of disciplines, countries, and universities. The Pakistan Research Rating and Benchmarking Tool is developed to rate and benchmark the progress of universities in a range of disciplines and sub-disciplines. Here, bibliometrics has been employed at a finer level than any other available evaluation system. In addition to disciplines like Engineering, sub-disciplines and niche areas like energy or architectural engineering are also dealt with. The tool will enable universities to assess their own strengths in disciplines and allocate resources and manpower optimally so as to invest in areas of research with promise for Pakistan. Most importantly, tool developed in this project will enable concerned personnel of universities and institutes working in science and technology in developing countries to indicate their particular niche strengths, benchmark and rate their quality of research, strengthen their individual research capabilities, and hence progress on their most suitable paths of development.

## Literature Review

### *The Structure of Global Research*

Phil Baty in his article on Areas of Expertise being pursued for world research stated, "Universities in East Asia are showing strength even in the humanities, a subject in which they traditionally have fared poorly [29]." The United States of America and the United Kingdom have always stood as the dominators of top echelons of ranking tables due to their patronage of giant educational and research institutes: Harvard, Oxford, and Cambridge Universities. However, in recent ranking tables, the US and its fellow "titans" in research and education are seeing their feet slipping as other countries muster serious efforts to join them in the race hence some of the major climbers have been from Asia [29].

The US remained in top ranks of education all across the globe but its breadth of placements in the ranking charts has been narrowing in the previous years [30]. For instance the combined ranking of Times Higher Education Supplement - Quacquarelli Symonds Limited in 2009 indicated institutions of Asian countries including Hong Kong, South Korea, Malaysia, and Japan to show up in considerably higher ranks of the charts. Phil Altbach associates this surge in performance of Asian universities in recent times because of increased investment by them in higher education. Also, these countries have promoted a hiring of international faculty to help improve their visibility as well as publishing of research in internationally recognized platforms.

### *Disciplines and Sub-Disciplines in Research*

Higher education and research institutes do not limit themselves to one field of knowledge. Instead a large range of subjects and specializations are pursued at every level. However, for ranking or benchmarking a university or research organization, optimal comparative data can be obtained if they are ranked as per their areas of expertise. On knowing where an institute stands within its various areas of specialization, more focused and appropriate plans of investment of resources can be made. An understanding of the subjects available i.e., disciplines, is necessary along with the various specializations i.e., sub-disciplines. Scopus, a popular bibliographic database, classifies research occurring across the globe in to four broad disciplinary: (i) Life Sciences, (ii) Physical Sciences, (iii) Health Sciences, and (iv) Social Sciences and Humanities [31]. Scopus further divides theses disciplines into 27 sub-disciplines and more than 300 sub-sub-disciplines or more precisely niche areas of research (Figure 1). Scopus categorizes research into these disciplines, sub-disciplines, and niche areas in order to categorize the journals it indexes.

The Institute of Scientific Information (ISI) uses its subject categories to classify all journals that are a part of the Science Citation Index (SCI) [32]. The solution that has been presented by ISI is also a way of categorizing science into its disciplines i.e., it can be interpreted as the *disciplinary structure of science*. ISI has developed 22 broad fields or disciplines in order to classify journals in. The University-Industry Research Connections (UIC) divides research into seven broad categories called *All Sciences* on the basis of which the popular Leiden Ranking is conducted [33]. These broad categories include cognitive sciences; earth and environmental sciences; life sciences; mathematics, computer sciences, and engineering; medical science; natural sciences; and social sciences. While research being conducted by a university may belong to only one of those broad fields, it can belong to several of the more specialized subject categories that are more fine-grained to fit the scope of journals.

### *Bibliometric indicators*

Since the initiation of bibliometrics in 1960s, bibliometricians and scientometricians have become more interested in devising ways of assessing publication impact [50]. Bibliometric practices have been changing over time and are expected to continue evolving. The practice of reliable bibliometrics are ensured by introducing standards or indicators: counting of papers released in association with countries, institutions, or researchers; counting of citations for work published earlier to measure the impact and need of a particular work; counting of co-citations; etc. [47]. These standards combine together to provide a more detailed and hence effective measurements to ensure the reliability of bibliometric results and methods adopted to produce them [49]. For standardized bibliometrics practices that can be used to measure research activity at any scientific level, we require adequate sources of documentation and data processing with a clear definition of all indicators. Glanzel emphasizes upon the fact that standards in bibliometrics have to be modified from time to time after being developed so as to satisfy the changing demands of science and technology [49].

In order to measure the scientific dynamism and orientation of a university, research institute, or country i.e., estimate its impact on international and national scientific and technology scenario, numerous relevant indicators or parameters have been introduced based on bibliometric analysis. Bibliometrics along with other measurement modes are used more and more to devise a systematic ways of ranking and benchmarking [50]. These indicators are indirect measures of the structure and output of a particular scientific community. While each indicator has its own advantages and limitations, bibliometrics provides a method of objective quantitative measure of scientific output. Some of the best known bibliometric indicators have been reviewed extensively by Yoshiko Okubo [47]. Some indicators with their advantages and disadvantages are discussed in Table 2

**Table 1** Advantages and disadvantages of certain significant bibliometric indicators (Adapted from [54])

| Bibliometric indicator | Advantages | Disadvantages |
|---|---|---|
| **Total number of papers ($N_p$)** | • It measures productivity. | • It does not measure importance or impact of papers. |
| **Total number of citations ($N_{c,tot}$)** | • It measures total impact. | • This indicator is hard to find and may be inflated by a small number of "big hits," which may not be representative of the individual if he or she is a coauthor with many others on those papers.<br>• It gives undue weight to highly cited review articles *vs.* original research contributions |
| **Citations per paper (i.e., ratio of $N_{c,tot}$ to $N_p$)** | • It allows comparison of scientists of different ages. | • It is hard to find.<br>• It rewards low productivity.<br>• It penalizes high productivity. |
| **Number of "significant papers, (The number of papers with >$y$ citations)** | • It eliminates the disadvantages of the above three criteria.<br>• It gives an idea of broad and sustained impact. | • $y$ is arbitrary and will randomly favor or disfavor individuals.<br>• $y$ needs to be adjusted for different levels of seniority. |
| **Number of citations to each of the $q$ most-cited papers** | • It overcomes many of the disadvantages of the criteria above | • It is not a single number, making it more difficult to obtain and compare.<br>• $q$ is arbitrary and will randomly favor and disfavor individuals. |

### *Benchmarking and Ranking Practice in Pakistan*

For Higher Education Commission (HEC) of Pakistan, enhancing the quality of education is a primary aim. HEC is struggling to bring Pakistani universities at par with the world demands by benchmarking them against the best within its own universities and also with internationally recognized universities [69]. Ranking is the most important mean that can help measure the effectiveness of all efforts that HEC puts to inculcate international competitiveness in research and academics of its higher education institutes. 2013, Higher Education Commission of Pakistan released the methodology through which the higher education institutes of the country are ranked according to their quality of education and research. The document discussed the methodology, criteria, and weightage give to various selected indicators [69].

In 2006, Pakistan became the first Islamic state to have presented a ranking for its higher education institutes through a properly designed ranking system [69]. According to the report of HEC in 2013 regarding its ranking system, "HEC believes that this ranking system will help policy making and visibility of the Pakistani universities as well." Ranking has always been a controversial task for the publishers but despite all the difficulties, Pakistan through its HEC released ranking twice: in 2006 followed by another ranking in 2011.

The latest ranking of higher education institutes in Pakistan by HEC was presented in 2015 at www.hesis.hec.gov.pk/stats and was based on data from years 2013-2014 [69]. This ranking included only those academic and research institutes that were recognized and registered by HEC and are ranked according to their area of specialization, which is decided as per the information provided by the institute itself.

It is important to know the research profiles of universities and research institutes as it sheds light upon their areas of strength including at discipline and niche areas. Knowledge of research profiles is bound to help in ideal investments by governments and other stakeholders in focused areas of research that are producing promising quality of research and hence are contributive to the laymen. Good quality research being conducted by a research group is bound to enhance the intellectual creative capacities at the level of individuals too. However, the challenge is to extend it to countries by helping them locate, adopt, and develop their most appropriate paths toward progress of science and technology. Hence the need of the time is that all parties involved, i.e., stakeholders, governments, universities, and research institutes, should be provided with tools able to help them better assess and develop their own research capacities in order to make the most out of whatever scarce resources are available.

Currently, no easily workable or accessible tools are available to locate active niche areas of research bodies in Pakistan. Furthermore, there are no means to measure the research performance of institutes in their niche areas through rating or benchmarking. While, the current global ranking systems including the Times Higher Education Ranking caters only 3 to 5% of the universities worldwide, which includes only the top 500 to 1000 universities of the world, the tool will enable universities to assess their own strengths in disciplines and benchmark and rate their quality of research as compared to other international standards.

**Methodology**

Data for the development was obtained from various sources in different steps. However, the primary data regarding universities and research institutes of Pakistan and their publications was obtained from Scopus. Scopus is the largest indexing service for journals with more than 14,000 titles in its database belonging to science and technology published from all over the world [27]. Keeping in mind its 100% coverage of journals in science and technology, which is the focus of this research thesis, we have utilized data from Scopus for development of tool.

Data for 222 Pakistani universities as well as research institutes was obtained from Scopus. The data included journal titles and publication titles. The number of publications downloaded from Scopus was 81,485 while the number of journals covered was 16,867.

### *Determining Subject Area of given Research Publication*

The subject area of every publication was determined through the All Journal Science Classification System (ASJC). Information based on SRC rank of journal, the ASJC name (discipline), and its MOD ASJC code was obtained and stored. The SRC rank of journal is a measure of the quality of research being published. In order to determine the exact subject area, ASJC was used again.

### *Subject Area Hierarchy in ASJC*

Information from ASJC was further obtained in the form of subject level hierarchy based on top level of subject i.e., discipline, the ASJC Level-1, which includes sub-disciplines, and the ASJC Level-2, which includes niche areas of research (Figure 7).

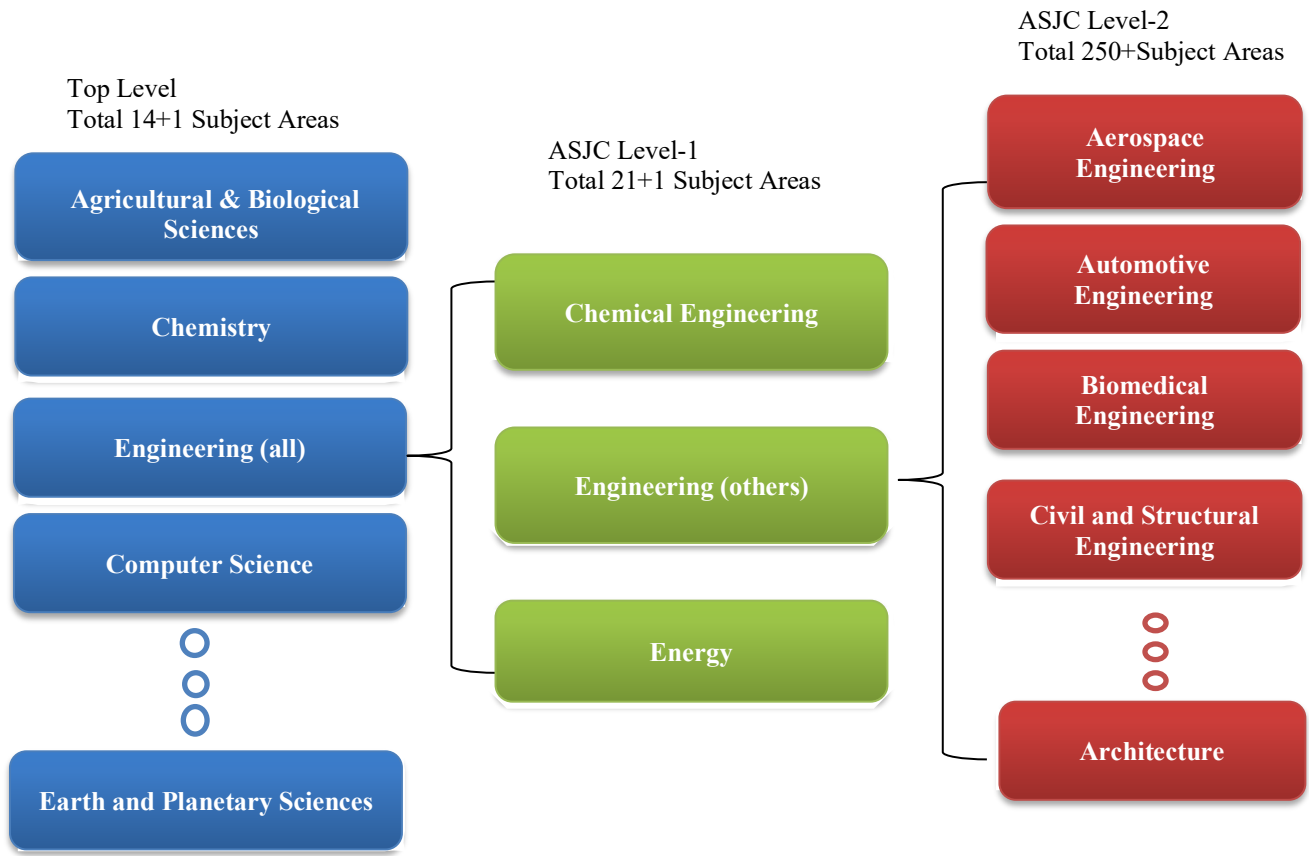

**Figure 1.1** Subject level hierarchy: Level 1 Discipline; Level 2 Sub-discipline; Level 3 Niche areas

***Subject Area Mapping with Scopus Data***

The data obtained from Scopus was combined with data gathered regarding ASJC subject categories (Figure 10). However, there was one issue in this regard i.e., one research journal could fall in two or more categories instead of just one. For example, *Vision Research* may be included in both Ophthalmology and Sensory Systems.

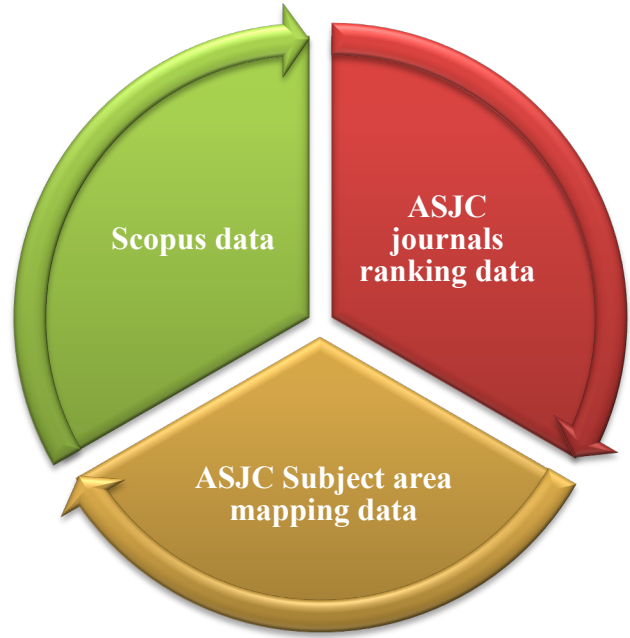

**Figure 1.2** Steps towards development of data for tool

***Computation of Bibliometric Indicators***

Collection of data for the tool was followed by the critical step of choosing bibliometric indicators. The bibliometric indicators of quality of research for any university and research organization were chosen and given different weightage according to the research aspect to be measured i.e., research volume or research quality. The tool estimates the volume of research conducted, quality of research, as well as and volume of research with good impact over the relevant field in the three services it provides.

- University rating
- University benchmarking

These bibliometric indicators in all of the aforementioned services are adjusted as follows:

***Volume-focused weightage***

In the volume-focused ranking of universities, the following 3 bibliometric indicators are given a weightage $\neq 0$ while rendering the remaining with a weight of 0. This has a default weight of 100 given to the following three indicators.

- *Total number of publications:* This is the sum of all publications for a particular university in a specific subject area.
- *Total number of citations:* This is the sum of the citations for a particular university in a specific subject area.
- *h*-index: *h*-index is the number of papers that have at least *h* citations each in a given subject area.

**Table 2** Bibliometric data for five chosen indicators used in the tool

| University | ASJC Code | Subject | ASJC Level | Total Pubs | % Pubs in 25% SNIP | Total Cites | H_INDEX | CPP |
|---|---|---|---|---|---|---|---|---|
| Lahore University of Management Sciences | 160000 | Physics and Astronomy (all) | 1 | 49 | 21 | 195 | 8 | 3.98 |
| University of Karachi | 11100 | Agricultural and Biological Sciences (others) | 2 | 722 | 63 | 1853 | 15 | 2.57 |
| University of Agriculture Faisalabad | 11100 | Agricultural and Biological Sciences (others) | 2 | 1664 | 58 | 7665 | 31 | 4.61 |
| Quaid-i-Azam University | 92310 | Pollution | 3 | 47 | 12 | 324 | 11 | 6.89 |
| Forman Christian College Lahore | 82209 | Industrial and Manufacturing Engineering | 3 | 5 | 2 | 7 | 2 | 1.4 |

### *Quality-focused weightage*

In quality-focused ranking of universities, the weightage assigned to bibliometric indicators is that the follow to indicators have weights ≠ 0, thus eliminating those that focus on quality. This has a default weight of 100 given to the following three indicators.

- %pubs in top 25% SNIP: These are the publications that are published in journals that are within top 25% of the subject area under consideration based on the SNIP value of the year 2010.
- Citation per paper (CPP): This bibliometric indicator is basically a ratio of total number of citations to total number of publications.

In this way, the final data obtained is shown in figure 11.

***All weights equal and custom weightage***

The tool is designed to provide users a balanced system of rating too. Users wishing to obtain the overall rating of universities irrespective of their greater production of research literature or better quality can use the "All weights equal" option at the tool. This has a default weight of 50 given to all five indicators.

## Development of framework for Rating of Universities

In subject specific rating, rating is provided from among more than 250 niche areas while in overall rating, rating of a given university is provided in multiple subject areas simultaneously in top 15 level 1 subjects of ASJC.

Several important selections to be made before a user can perform rating are: range of years (2008-2013), region (all, provincial), subject area (level 1), subject area (level 2), subject area (level 3) and indicator weightage.

In order to rate universities as per the subject, the tool performs several steps starting from assigning weights till band allocation.

***Normalization of data***

First of all, the tool works by identifying the maximum numerical value for computed indicators. This maximum value is used to normalize the others by dividing them with it. The numerical values are already inserted in the program in forms. This step provides the normalized indicator values of the five indicators for each university in a given subject area.

***Assigning weights to indicators***

Moving on, the normalized values of the bibliometric indicators are multiplied with the assigned weight (0-100) of that indicator. The products thus obtained are the weighted normalized indicator values.

***Scoring and percentage score***

The weighted normalized indicator values are summed for each university and a grand total is obtained. In the next step, the tool again indicates the highest grand total and uses it to divide every other grand total. Subsequently, these are multiplied to 100 to obtained the percentage score for each university in a given subject area. This score varies from 0 to 100.

*Band Allocation to universities*

The tool is then designed to convert these percentages to bands: the band allocated to each university is its final score indicating its performance in that subject. A score of 91-100 translates to allocation of band 1, which indicates a high overall performance in that subject area. Remaining band allocation is shown in table.

**Table 3** Band allocation according to percentage

| Percentage score | Band |
|---|---|
| **91-100** | 1 |
| **81-90** | 2 |
| **71-80** | 3 |
| **61-70** | 4 |
| **51-60** | 5 |
| **41-50** | 6 |
| **31-40** | 7 |
| **21-30** | 8 |
| **11-20** | 9 |
| **1-10** | 10 |

**Algorithm working behind "university rating" service of the tool**

In the former, the above process till band allocation is conducted for chosen niche areas in a discipline with every university falling in the domain being assigned a band representing its performance. However, in the latter, all universities in a chosen region are rated within the top 15 AJSC subject areas level 1 subjects given in (subject level hierarchy one). Here the bands are pre-calculated for every university in the 15 level 1 subjects in three scenarios (i) All eights equal, (ii) quality focused, and (iii) volume focused as shown in Figure of algo one.

*Subject Specific Rating*

In subject specific rating the tool provide facility to rate universities in 255 niche subject area based on Volume-focused weightage, Quality-focused weightage and All Weights Equal weightage. There is facility for the user to custom the weights of all the indicators according to the need to research demographics of that region.

**Development of framework for benchmarking of universities**

The second major service provided by the tool is benchmarking. For benchmarking service, major selections required to be made by the users include: range of years (2008-2013), region (all Pakistan, provincial), subject area (level 1), subject area (level 2), and subject area (level 2). Benchmarking works with a tool designed to yield results in the form of either spider-web chart if "at a glance" benchmarking is chosen or as bar and scatter chart if "detailed benchmarking" is required. In either case, the tool is designed to allow cross-university analysis. This has been achieved as follows:

***Benchmarking a glance***

This section is designed for cross comparison of different universities in a particular subject area along different indicators. User can also cross examine the universities on multiple subject areas by using multiple tab options.

***Normalization of bibliometric indicators***

The algorithm is designed to at first normalize all five indicators for the chosen universities (that can be up to 5) after selection of the aforementioned criteria. Normalization of each bibliometric indicator is performed by dividing all numerical values with the highest value for that indicator among the chosen universities.

***Percentage scores for bibliometric indicators***

All the resultant values are then multiplied with 100 to get the percentage.

***Plotting on spider-web chart***

These percentage value are then plotted on spider-web chart (showing actual values on labels) for the cross comparison of different universities in a particular subject area along different indicators. User can also cross examine the universities on multiple subject areas by using multiple tab options.

## Results and Discussions

### University Subject Specific Rating Analysis

In this section universities are rated according to specific subject area. All the indicators are normalized according to their respective subject field as discussed in the previous section of methodology.

We are presenting case study of few subject areas of different ASJC level based on assigning equal weightages to all the indicators, volume focused and quality focused.

Talking about subject area ASJC level 1 ***Engineering All*** where all weights are equal and min publication threshold is forty, we get the following results from the tool.

**Table 4** University Subject Specific Rating Analysis subject area ASJC level 1 Engineering All where all weights are equal and min publication threshold is forty

| University | Publication | Citation | H-index | % Pubs in top 25% SNIP | CPP | Band |
|---|---|---|---|---|---|---|
| **Quaid-i-Azam University** | 905 | 6224 | 32 | 42 | 6.88 | 1 |
| **COMSATS Institute of Information Technology** | 725 | 3029 | 24 | 36 | 4.18 | 4 |
| **University of Agriculture Faisalabad** | 173 | 2123 | 24 | 38 | 12.27 | 4 |
| **National University of Sciences and Technology Pakistan** | 636 | 2483 | 23 | 34 | 3.9 | 4 |
| **Pakistan Institute of Nuclear Science and Technology** | 278 | 1758 | 19 | 49 | 6.32 | 5 |
| **Pakistan Institute of Engineering and Applied Sciences** | 257 | 1051 | 14 | 50 | 4.09 | 6 |
| **Bahauddin Zakariya University** | 140 | 726 | 15 | 47 | 5.19 | 6 |
| **University of Sindh** | 84 | 540 | 14 | 43 | 6.43 | 6 |
| **University of the Punjab Lahore** | 257 | 1047 | 15 | 38 | 4.07 | 6 |
| **University of Peshawar** | 167 | 612 | 13 | 37 | 3.66 | 7 |

Talking about subject area ASJC level 1 ***Engineering All*** where volume focused indicators are assigned by 100% weightage and others at 0 and min publication threshold is forty, we get the following results from the tool.

**Table 5** University Subject Specific Rating Analysis where subject area ASJC level 1 Engineering All where volume focused indicators are assigned by 100% weightage and others at 0 and min publication threshold is forty

| University | Publication | Citation | H-index | % Pubs in top 25% SNIP | CPP | Band |
|---|---|---|---|---|---|---|
| **Quaid-i-Azam University** | 905 | 6224 | 32 | 42 | 6.88 | 1 |
| **COMSATS Institute of Information Technology** | 725 | 3029 | 24 | 36 | 4.18 | 4 |
| **National University of Sciences and Technology Pakistan** | 636 | 2483 | 23 | 34 | 3.9 | 4 |
| **University of Agriculture Faisalabad** | 173 | 2123 | 24 | 38 | 12.27 | 6 |
| **Pakistan Institute of Nuclear Science and Technology** | 278 | 1758 | 19 | 49 | 6.32 | 7 |
| **University of the Punjab Lahore** | 257 | 1047 | 15 | 38 | 4.07 | 7 |
| **University of Peshawar** | 167 | 612 | 13 | 37 | 3.66 | 8 |
| **Pakistan Institute of Engineering and Applied Sciences** | 257 | 1051 | 14 | 50 | 4.09 | 8 |
| **Bahauddin Zakariya University** | 140 | 726 | 15 | 47 | 5.19 | 8 |
| **University of Karachi** | 150 | 521 | 13 | 24 | 3.47 | 8 |

Talking about subject area ASJC level 1 ***Engineering All*** where quality focused indicators are assigned by 100% weightage and others at 0 and min publication threshold is forty, we get the following results from the tool.

**Table 6** University Subject Specific Rating Analysis where subject area ASJC level 1 Engineering All where quality focused indicators are assigned by 100% weightage and others at 0 and min publication threshold is forty

| University | Publication | Citation | H-index | % Pubs in top 25% SNIP | CPP | Band |
|---|---|---|---|---|---|---|
| **University of Agriculture Faisalabad** | 173 | 2123 | 24 | 38 | 12.27 | 1 |
| **Pakistan Institute of Nuclear Science and Technology** | 278 | 1758 | 19 | 49 | 6.32 | 3 |
| **National University of Computer and Emerging Sciences Islamabad** | 57 | 246 | 9 | 60 | 4.32 | 3 |
| **Quaid-i-Azam University** | 905 | 6224 | 32 | 42 | 6.88 | 4 |
| **Lahore University of Management Sciences** | 53 | 128 | 6 | 64 | 2.42 | 4 |
| **Pakistan Institute of Engineering and Applied Sciences** | 257 | 1051 | 14 | 50 | 4.09 | 4 |
| **Bahauddin Zakariya University** | 140 | 726 | 15 | 47 | 5.19 | 4 |
| **University of Sindh** | 84 | 540 | 14 | 43 | 6.43 | 4 |
| **Hazara University Pakistan** | 57 | 197 | 9 | 40 | 3.46 | 5 |
| **University of Sargodha** | 75 | 407 | 12 | 33 | 5.43 | 5 |

The above three tables shows the results of university rating in engineering all ASJC level 1. We can see variance very clearly in the results. If we talk about Quaid-i-Azam University we can see that it has band 1 in case of all weight equal and volume focused but looking at quality focused it has band 4; this means that Quaid-i-Azam University is focusing on expanding out and less focused on expanding up. Similarly if we talk about National University of Computer and Emerging Sciences Islamabad and Lahore University of Management Sciences they are not existing in the top 10 list of of all weight equal and volume focused indicators but if we look at quality focused indicators they have band 3 and band 4 respectively; this means that even though they are producing less publications but their majority publication are of good quality.

We can also see the change in the bands of different universities against different weightage of indicators. This clearly shows that it is not necessary that a university producing more research publication is always producing well in quality and vice versa.

Talking about subject area ASJC level 3 ***Electrical and Electronic Engineering*** where all weights are equal and min publication threshold is forty, we get the following results from the tool.

**Table 7** University Subject Specific Rating Analysis where subject area ASJC level 3 Electrical and Electronic Engineering where all weights are equal and min publication threshold is forty

| University | Publication | Citation | H-index | % Pubs in top 25% SNIP | CPP | Band |
|---|---|---|---|---|---|---|
| **Quaid-i-Azam University** | 171 | 1201 | 18 | 31 | 7.02 | 1 |
| **COMSATS Institute of Information Technology** | 197 | 575 | 12 | 30 | 2.92 | 4 |
| **National University of Sciences and Technology Pakistan** | 170 | 518 | 11 | 33 | 3.05 | 4 |
| **Pakistan Institute of Nuclear Science and Technology** | 22 | 168 | 9 | 45 | 7.64 | 5 |
| **Center for Advanced Studies in Engineering Pakistan** | 22 | 104 | 6 | 55 | 4.73 | 6 |
| **Ghulam Ishaq Khan Institute of Engineering Sciences and Technology** | 69 | 280 | 9 | 25 | 4.06 | 6 |
| **National University of Computer and Emerging Sciences Islamabad** | 21 | 95 | 7 | 76 | 4.52 | 6 |
| **Mohammad Ali Jinnah University** | 22 | 85 | 5 | 45 | 3.86 | 7 |
| **University of the Punjab Lahore** | 30 | 134 | 7 | 20 | 4.47 | 7 |
| **International Islamic University Islamabad** | 22 | 60 | 5 | 14 | 2.73 | 8 |

**Table 8** University Subject Specific Rating Analysis where subject area ASJC level 3 Electrical and Electronic Engineering where volume focused indicators are assigned by 100% weightage and others at 0 and min publication threshold is forty

| University | Publication | Citation | H-index | % Pubs in top 25% SNIP | CPP | Band |
|---|---|---|---|---|---|---|
| Quaid-i-Azam University | 171 | 1201 | 18 | 31 | 7.02 | 1 |
| COMSATS Institute of Information Technology | 197 | 575 | 12 | 30 | 2.92 | 3 |
| National University of Sciences and Technology Pakistan | 170 | 518 | 11 | 33 | 3.05 | 4 |
| Ghulam Ishaq Khan Institute of Engineering Sciences and Technology | 69 | 280 | 9 | 25 | 4.06 | 7 |
| Pakistan Institute of Engineering and Applied Sciences | 53 | 131 | 7 | 19 | 2.47 | 8 |
| Pakistan Institute of Nuclear Science and Technology | 22 | 168 | 9 | 45 | 7.64 | 8 |
| University of the Punjab Lahore | 30 | 134 | 7 | 20 | 4.47 | 8 |
| International Islamic University Islamabad | 22 | 60 | 5 | 14 | 2.73 | 9 |
| NED University of Engineering and Technology Pakistan | 22 | 75 | 6 | 18 | 3.41 | 9 |
| Bahauddin Zakariya University | 20 | 78 | 6 | 0 | 3.9 | 9 |

Talking about subject area ASJC level 3 ***Electrical and Electronic Engineering*** where volume focused indicators are assigned by 100% weightage and others at 0 and min publication threshold is forty, we get the following results from the tool. The above two tables show the results of university rating in ASJC level 3 Electrical and Electronic Engineering. We can see that COMSATS Institute of Information Technology is rising up one band from 4 to 3 looking from all weights equal table to volume focused table respectively. These result also shows the performance of universities among niche subject areas which was not analyzed before.

**Cross University benchmarking Analysis**

Talking about **Computer Science all** ASJC Level 1 in five Public Private Sector large Universities of Pakistan we look at the benchmarking of Quaid-i-Azam University, Lahore University of Management Sciences, COMSATS Institute of Information Technology, National University of Sciences and Technology Pakistan and National University of Computer and Emerging Sciences Islamabad.

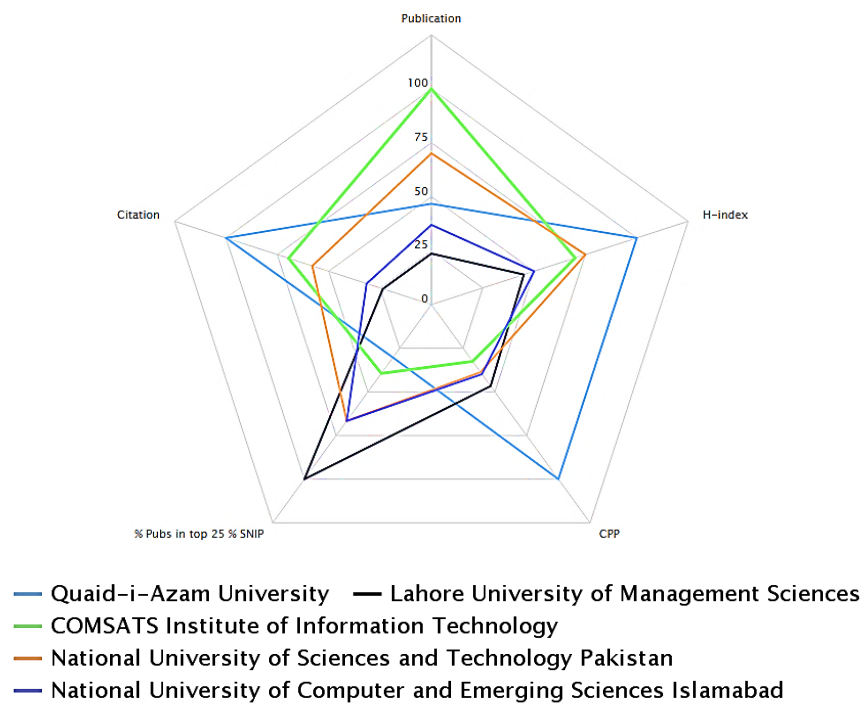

**Figure 1.3** benchmarking at glance of five Public Private Universities of Pakistan in Computer Science all

Talking about the *figure 4.22* it shows benchmarking of five Public Private Universities of Pakistan in Computer Science. The results indicates that COMSATS Institute of Information Technology and National University of Sciences and Technology Pakistan are producing more research in this field but they producing less quality work in comparison to others. Numbers of publications of these two universities are higher but the other entire indicators are not so much attractive. On the other hand Quaid-i-Azam University is producing research almost half of the volume of COMSATS Institute of Information Technology, but three indicators i.e. cpp, h-index and citations are much higher than above two universities. Lahore University of Management Sciences is not performing that much good in all the other indicators but it's percentage of publication published in top 25% SNIP is much higher than all of the other institutes. This clearly shows that LUMS is publishing its work in top journals in the field of computer science.

If we benchmark the above five universities we can say that Quaid-i-Azam University has good benchmarking in volume based and quality indicators thus focused on expand out and expand up in research and Lahore University of Management Sciences has good benchmarking in quality based indicator thus focused on expand up in research. The other two COMSATS Institute of Information Technology and National University of Sciences and Technology are performing average in overall results of this subject area. Talking about Computer Science (all) -> Computer Science (all) -> **Artificial Intelligence** ASJC Level 3 in five Public Private Sector large Universities of Pakistan we look at the benchmarking of Quaid-i-Azam University, Lahore University of Management Sciences, COMSATS Institute of Information Technology, National University of Sciences and Technology Pakistan and National University of Computer and Emerging Sciences Islamabad.

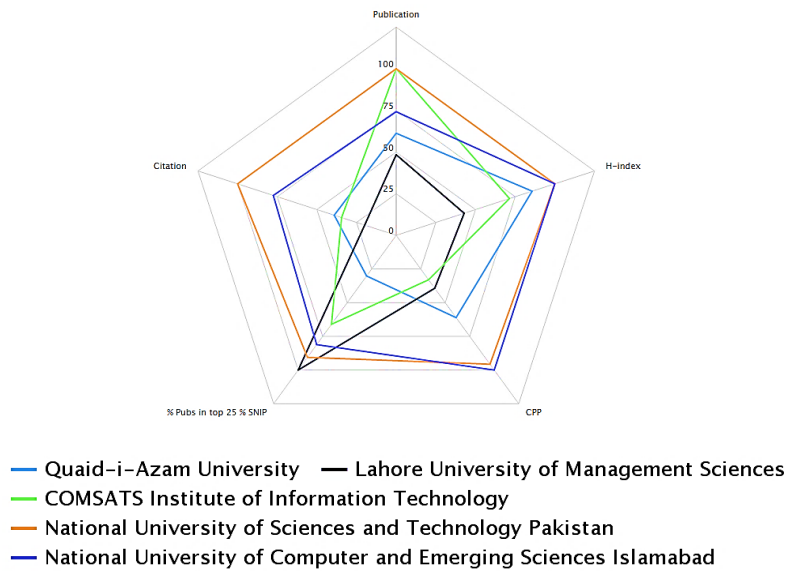

**Figure 1.4** benchmarking at glance of five Public Private Universities of Pakistan in Artificial Intelligence

Talking about the *figure 4.28* it shows benchmarking of five Public Private Universities of Pakistan in Artificial Intelligence. The results indicates that COMSATS Institute of Information Technology and National University of Sciences and Technology Pakistan are producing more research in this field but they producing less quality work in comparison to others. Numbers of publications of these two universities are higher but the other entire indicators are not so much attractive. On the other hand Quaid-i-Azam University is producing research almost half of the volume of COMSATS Institute of Information Technology, but three indicators i.e. cpp, h-index and citations are much higher than above two universities. Lahore University of Management Sciences is not performing that much good in all the other indicators but it's percentage of publication published in top 25% SNIP is much higher than all of the other institutes. This clearly shows that LUMS is publishing its work in top journals in the field of computer science.

If we benchmark the above five universities we can say that Quaid-i-Azam University has good benchmarking in volume based and quality indicators thus focused on expand out and expand up in research and Lahore University of Management Sciences has good benchmarking in quality based indicator thus focused on expand up in research. The other two COMSATS Institute of Information Technology and National University of Sciences and Technology are performing average in overall results of this subject area.

Talking about Engineering (all) -> Energy (all) -> ***Nuclear Energy and Engineering*** ASJC Level 3 in four Public Private Sector small Institutes and one large University of Pakistan we look at the benchmarking of Quaid-i-Azam University, University of Peshawar, Pakistan Atomic Energy Commission, Pakistan Institute of Engineering and Applied Sciences and Pakistan Institute of Nuclear Science and Technology.

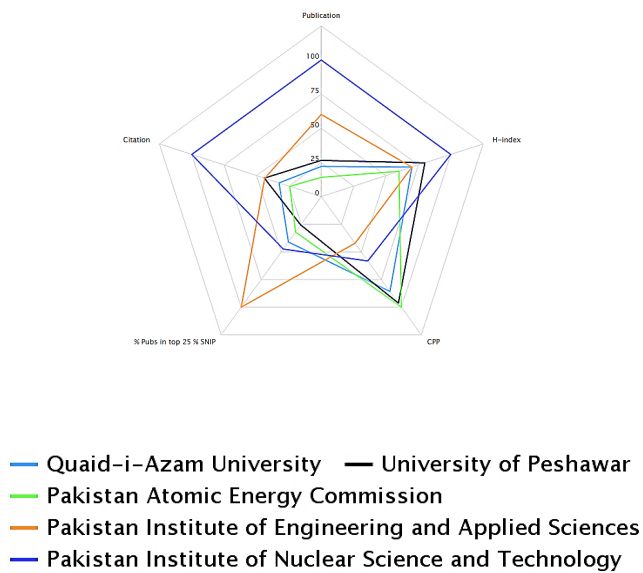

**Figure 1.5** benchmarking at glance of five Public Private Universities of Pakistan in Nuclear Energy and Engineering

Talking about Engineering (all) -> Energy (all) -> ***Nuclear Energy and Engineering*** ASJC Level 3 in four Public Private Sector small Institutes and one large University of Pakistan we look at the benchmarking of Quaid-i-Azam University, University of Peshawar, Pakistan Atomic

Energy Commission, Pakistan Institute of Engineering and Applied Sciences and Pakistan Institute of Nuclear Science and Technology.

Talking about the *figure 4.34* it shows benchmarking of four Public Private Sector small Institutes and one large University of Pakistan in Nuclear Energy and Engineering. The results indicate that of Quaid-i-Azam University is producing more research in this field but it is producing less quality work in comparison to others. On the other hand Pakistan Institute of Engineering and Applied Sciences is producing research almost half of the volume of Quaid-i-Azam University, but its cpp and percentage of publication published in top 25% SNIP are much higher than Quaid-i-Azam University. University of Peshawar, Pakistan Atomic Energy Commission and Institute of Nuclear Science and Technology are not performing that much good in all the other indicators but their cpp is much higher than all of the other institutes

If we benchmark the above five universities we can say that Quaid-i-Azam University has good benchmarking in volume based indicators thus focused on expand out in research and Pakistan Institute of Engineering and Applied Sciences has good benchmarking in quality based indicator thus focused on expand up in research.

**Conclusion and Summary**

In this research we present the research landscape of Pakistan; volume vs quality. Our main focus was to develop computational tool for the developing countries that will enable them to assess research performance of universities and research institutes through rating and benchmarking against standards or pre-determined indicators so as to adopt and invest in the best paths toward progress in science and technology. It also enables universities and research institutes to identify and evaluate their niche areas of expertise for the purpose of locating gaps in their research activities and develop themselves in areas of focus.

The tool has been developed with a clear goal in mind to cater universities as well as institutes/government agencies involved in research in its two defined functionalities: benchmarking and rating. Rating and benchmarking is provided in the disciplines of science and technology down to a wide range of specific niche areas defined in the All Journals Science Classification (ASJC) within a time frame of 2008-13. This computational tool is capable of measuring the volume and quantity of research in science and technology being conducted by universities and institutes of Pakistan.  In order to measure the scientific dynamism and orientation of a university, research institute, or country five most significant indicators are used for the development of the framework [70-74].

By virtue of the strong results our work makes an important contribution towards exploring the research landscape of Pakistan. We have analyzed some of the best public and private sector universities and found that the overall rating is not enough to measure the excellence of a university, one needs to identify the best niche subject areas of a University to determine its strong areas of research. Similarly, it is not always a fact that a large university is performing well in every subject area; there are several small universities which are producing quality work in many new cutting edge science and technology subjects.

Government and funds allocating bodies should take this factor in account that investing the right money at right place will give far better results than we they are having right now. Our tool has a capability to help them to identify the best institutes in a particular subject area in which they are interested to invest.

Currently we are dealing with the research dataset of Pakistan from the year 2008 to 2013. This will be enhanced to the previous missing years and upcoming years to provide the historical and future analytics of research quality and quantity. The tool is quiet flexible to support unlimited time windows of data. All we need to provide the dataset drawn from Scopus.